\documentclass{ws-ijmpa}
\usepackage{graphicx}
\usepackage[utf8]{inputenc}
\usepackage{t1enc}
\usepackage[english]{babel}
\usepackage{comment}
\usepackage{footnotebackref}
\usepackage{hyperref}
\usepackage{enumitem}
\usepackage{amssymb}
\usepackage{tabularx}
\usepackage{amsmath}
\usepackage{xcolor}
\usepackage[backend=biber,style=numeric,sorting=none]{biblatex}
\begin{document}
\markboth{G\'abor Kasza}{Describing the Thermal Radiation by Relativistic Hydrodynamics}

%
\catchline{}{}{}{}{}
%

\title{Describing the Thermal Radiation in $Au+Au$ Collisions at $\sqrt{s_{NN}}=$200\nolinebreak[4] GeV by an Analytic Solution of Relativistic Hydrodynamics}

\author{G\'abor Kasza}
\address{HUN-REN Wigner Research Centre for Physics, Budapest H-1121, Hungary\\
Hungarian University of Agriculture and Life Sciences, Gy\"ongy\"os H-3200, Hungary\\ 
kasza.gabor@wigner.hun-ren.hu}

\maketitle

\begin{history}
\received{Day Month Year}
\revised{Day Month Year}
\end{history}

\begin{abstract}
In high-energy heavy-ion collisions a nearly perfect fluid, the so-called strongly coupled quark gluon plasma forms. After the short period of thermalisation, the evolution of this medium can be described by the laws of relativistic hydrodynamics. The time evolution of the quark gluon plasma can be understood through direct photon spectra measurements, which are sensitive to the entire period between the thermalisation and the freeze-out of the medium. I present a new analytic formula that describes the thermal photon radiation and it is derived from an exact and finite solution of relativistic hydrodynamics with accelerating velocity field. Then I compare my calculations to the most recent nonprompt spectrum of direct photons for $Au+Au$ at $\sqrt{s_{NN}}=$200 GeV collisions. I have found a convincing agreement between the model and the data, which allows to give an estimate of the initial temperature in the center of the fireball. My results predict hydrodynamic scaling behaviour for the thermal photon spectra of high-energy heavy-ion collisions.

\end{abstract}

\ccode{PACS numbers:}

\section{Introduction}

According to the current view of modern physics, the Universe around us was created at the moment of the Big Bang, and from there we understand the existence of space-time. One of the main tasks of high-energy heavy-ion physics is to reveal the state of the fractions of a millisecond after the moment of creation. At this time, the Universe was hot and pressurised to an unimaginable degree in ordinary terms, so that an unusual matter, made up of quarks and gluons, filled the space. This medium is called quark gluon plasma (QGP). The theory of quarks and gluons is summarised in quantum colour dynamics (QCD). However, in the first half of the 2000s, four experiments of the RHIC accelerator~\cite{BRAHMS:2004adc,PHENIX:2004vcz,PHOBOS:2004zne,STAR:2005gfr} discovered that many of the properties of QGP cannot be described in the perturbative discussion of QCD, since in this medium the collision cross section of quarks diverges and the mean free path is close to zero rather than infinity. Consequently, the QGP does not behave as an ideal gas but as a near-perfect quark fluid~\cite{Lacey:2006bc}.

The equations of hydrodynamics have no internal physical scale, so they can be applied from the smallest, experimentally achievable scales, to the largest, cosmological distances. As a result, hydrodynamic equations can be used to describe the time evolution of our Universe starting from the Big Bang, but they can also be used to study the time evolution at the smallest femtometre distances, where the "Little Bangs" of high-energy heavy-ion collisions also create fireballs evolving according to the rules of hydrodynamics. Consequently, the solutions of relativistic hydrodynamics are excellent candidates for describing observables in high-energy heavy-ion collisions.

The hadron spectra obtained from the different experiments reflect the moment of the freeze-out, at which point the hadrons fly towards the detectors without interaction. Thus, the hadronic observables are independent of the initial state of the QGP or the equation of state characterising the microscopic properties of the medium. As a result, the hadronic observables do not provide detailed information on the time evolution of the properties of the QGP before the freeze-out. However, shortly after the medium is created, the system is thermalised, and from then until the freeze-out, direct photons are produced directly from the quark matter. These photons can penetrate the medium and reach the detectors without interaction due to the small cross section of the electromagnetic interaction and the fact that the photons do not participate in the strong interaction. Since the direct photons traverse the medium unmodified, they encode information about the environment, such as the temperature or the collective motion. Hence, the direct photon spectrum is an excellent probe for understanding the time evolution of the temperature of the QGP.

The high transverse momentum ($p_{\rm T}$) regime of the direct photon spectra is dominated by the photons emitted in high scattering processes, but the low-$p_{\rm T}$ regime can be mostly considered the contribution of the thermal radiation (other unconventional sources are also possible, see for example Refs.~\cite{Basar:2012bp,Basar:2014swa,Gale:2021emg,Linnyk:2015tha}). Therefore, only the low-$p_{\rm T}$ regime can be described by hydrodynamic models, or those spectra from which the hard scattering contribution has been subtracted (these latter are called nonprompt direct photon spectra).

In this paper, I extract an analytic formula for the thermal contribution to the direct photon spectrum from a recently found solution of relativistic hydrodynamics~\cite{Csorgo:2018pxh}. Then, I compare this analytic formula with a recently published dataset of the PHENIX collaboration~\cite{PHENIX:2022rsx}, which describes the nonprompt direct photon spectrum measured in $Au+Au$ collisions at $\sqrt{s_{NN}}=$200 GeV. Similar efforts was successfully done in Refs.~\cite{Csanad:2009wc,Csanad:2011jq}, where the authors used a 1+3 dimensional, Hubble-type solution of relativistic hydrodynamics to describe the direct photon spectrum and hadronic observables for PHENIX $Au+Au$ collisions at $\sqrt{s_{NN}}=$200 GeV.

\section{The Analyzed Solution of Relativistic Hydrodynamics}\label{sec:CKCJ}

In Ref.~\cite{Csorgo:2018pxh}, a new family of 1+1 dimensional parametric solutions of relativistic perfect fluid hydrodynamics has been found. In this family of solutions, the equation of state is defined by the following expressions together:
\begin{align}
    \varepsilon &= \kappa p,\label{eq:eos}\\
    \mu &= 0,
\end{align}
where $\varepsilon$ is the energy density, $p$ stands for the pressure, the chemical potential is denoted by $\mu$ and $\kappa$ is a constant and relates to the speed of sound: $\kappa=c_s^{-2}$. The velocity field characterizes a locally accelerating fluid and it is written as:
\begin{align}
    u^{\mu}&=\left(\cosh\left(\Omega\right),\sinh\left(\Omega\right)\right),\label{eq:velocity_u}\\
    \Omega\left(\eta_z\right) &= \frac{\lambda}{\sqrt{\lambda-1}\sqrt{\kappa-\lambda}}\arctan\left(\sqrt{\frac{\kappa-\lambda}{\lambda-1}}\tanh\left(\Omega(\eta_z)-\eta_z\right)\right),\label{eq:velocity_O}
\end{align}
where $\eta_z$ is the space-time rapidity and $\Omega$ stands for the fluid rapidity, which can be given only in implicit form and depends only on the space-time rapidity. The rate of acceleration is denoted by $\lambda$. In the boost-invariant Hwa-Bjorken solution~\cite{Hwa:1974gn,Bjorken:1982qr}, this $\lambda$ parameter is 1, so the fluid rapidity and the space-time rapidity are equal to each other: $\Omega=\eta_z$. Such a flow profile leads to unrealistic, flat multiplicity distributions. The solutions of Ref.~\cite{Csorgo:2018pxh} have been found with locally accelerating velocity field, expressed by Eq.~\eqref{eq:velocity_u} and Eq.~\eqref{eq:velocity_O}, which makes them ideal candidates to describe the thermal photon spectrum. 

In Ref.~\cite{Csorgo:2018pxh}, not only one solution but a whole family of solutions was presented, since the solution for the temperature field can be multiplied by any arbitrary function of the scale variable $s$: let me denote it by $\mathcal{T}(s)$. The scale variable $s$ is given by the $u^{\mu}\partial_{\mu}s=0$ scale equation and it determines the trajectory of the fluid elements. In the present paper, we select the simplest option from this family of solutions by fixing the $\mathcal{T}(s)$ scale function to 1. In this case, the temperature field of the expanding fireball is given by the following formula:
\begin{equation}\label{eq:temp}
    T\left(\tau,\eta_z\right) = T_0 \left(\frac{\tau_0}{\tau}\right)^{\frac{\lambda}{\kappa}}\left[1+\frac{\kappa-1}{\lambda-1}\sinh^2\left(\Omega\left(\eta_z\right)-\eta_z\right)\right]^{-\frac{\lambda}{2\kappa}},
\end{equation}
where $\tau$ is the longitudinal proper time, while $T_0$ and $\tau_0$ stand for the initial conditions. Considering that the chemical potential is 0, the other thermodynamic variables can be determined through the equation of state and the temperature field.

Although the solution we are discussing can only be interpreted on a certain interval of space-time rapidity ($\eta_z \in \left[-\eta_z^{max},\eta_z^{max}\right]$), in the $\lambda\rightarrow 1$ limit $\eta_z^{max}$ goes to infinity. This solution has been successfully used before to describe the pseudorapidity distribution of charged hadrons~\cite{Csorgo:2018pxh,Csorgo:2018fbz,Kasza:2018jtu,Kasza:2018qah}, and it was found that $\lambda$ is not much larger than 1. Based on this, I calculated the thermal photon spectrum in the $\lambda\rightarrow 1$ limit, so we do not have to worry about the finite nature of the solution.

Let me note, that in the $\kappa=1$ case, this solution reproduces the 1+1 dimensional solution of Refs.~\cite{Csorgo:2006ax,Nagy:2007xn}, while in the $\lambda\rightarrow 1$ limit the Hwa-Bjorken solution can be recovered~\cite{Hwa:1974gn,Bjorken:1982qr}, but to see the latter, one has to proceed carefully in the calculations (for more details, see Ref.~\cite{Csorgo:2018pxh}).

\section{New Analytic Formula for the Thermal Radiation}\label{sec:analytic_thermal_radiation}

In this chapter, I briefly describe the assumptions I used to evaluate the thermal photon spectrum. I have described the locally thermalized medium using the solution presented in the Sec.~\ref{sec:CKCJ} and published in Ref.~\cite{Csorgo:2018pxh}. Both the hadronic observables and the thermal radiation spectra can be derived from a source function describing the phase space distribution, but for hadrons the source function can only reflect the final state at the kinetic freeze-out. In contrast, the source function for thermal photons is sensitive to the whole time evolution of the fireball, and suppose that the QGP is thermalized by the strong interaction. Accordingly, the source function can be written as follows:
\begin{equation}\label{eq:source}
    S\left(x^{\mu},p^{\mu}\right)d^4 x=\frac{g}{\left(2\pi \hbar\right)^3} \frac{ H(\tau)}{\tau_{\rm R}} \frac{p_{\mu}d\Sigma^{\mu}}{\exp\left(\frac{p^{\mu}u_{\mu}}{T}\right)-1},
\end{equation}
where $g$ is the degeneracy factor, $\tau_{\rm R}$ stands for the characteristic time for radiation, the four-momentum of the photons is denoted by $p^{\mu}$ and $d\Sigma^{\mu}$ is the normal vector of the freeze-out hypersurface, which is assumed to be proportional to the velocity field of the expanding medium. The $p_{\mu}d\Sigma^{\mu}$ term is called Cooper-Frye factor, which relates to the flux of the particles.~\cite{Cooper:1974mv}. If the medium is described by the solution of Sec.~\ref{sec:CKCJ}, then $d\Sigma^{\mu}$ can be expressed as follows:
\begin{equation}
    d\Sigma^{\mu} = \frac{u^{\mu}\tau d\tau d\eta_z dr_x dr_y}{\cosh\left(\Omega\left(\eta_z\right)-\eta_z\right)},
\end{equation}
where the transverse plane ($r_x$, $r_y$) is also included, so the normal vector of the freeze-out hypersurface is embedded to the 1+3 dimensional space-time.  This implies that I assume that the temperature is homogeneous in the transverse plane. The  $H\left(\tau\right)$ function stands for the proper time distribution of the opacity of the QGP to photons. However, the mean free path of photons is larger than the size of the thermalized fireball, which means that the medium is transparent to photons. Thus, the $H(\tau)$ function simply describes the duration of thermal photon emission in proper time:
\begin{equation}
    H(\tau)=\Theta\left(\tau-\tau_{\rm f}\right)-\Theta\left(\tau-\tau_0\right),
\end{equation}
where $\tau_{\rm f}$ is the proper time at the freeze-out, $\tau_0$ stands for the initial proper time and $\Theta(\tau)$ denotes the Heaviside step function. Note that I neglected those photons that are emitted in the free-stream of hadrons after $\tau_{\rm f}$. The invariant momentum distribution of thermal radiation can be obtained by integrating the source function over space and time. I applied the Boltzmann approximation of the source function: the difference between the Bose-Einstein and Boltzmann distributions becomes significant in the low-$p_{\rm T}$ regime ($p_{\rm T}<1$ GeV), where, as far as I know, the thermal radiation of high-energy collisions has not yet been entirely mapped. I took advantage of the fact that I can also use the $\Omega\approx \lambda \eta_z$ approximation in the range close to midrapidity. This approximation corresponds to the $\lambda\rightarrow 1$ limit and it describes the $\Omega\left(\eta_z\right)$ relation well almost throughout the range of validity of the solution. The deviation from the linear trend occurs only very close to the edges of validity, which is negligible compared to the full range. In addition, I used saddle-point approximation to perform the integrals. The detailed derivation of the thermal photon spectrum at midrapidity can be found in the Appendix. In this section, I only report the result, which is expressed as follows:
\begin{equation}\label{eq:thermal_radiation_result}
    \left.\frac{d^2 N}{2\pi p_{\rm T} dp_{\rm T} dy}\right|_{y=0}=N_0\:\frac{2\alpha}{3\pi^{3/2}}\left[\frac{1}{T_{\rm f}^{\alpha}}-\frac{1}{T_{0}^{\alpha}}\right]^{-1} p_{\rm T}^{-\alpha-2}\left.\Gamma\left(\alpha+\frac{5}{2},\frac{p_{\rm T}}{T}\right)\right|^{T=T_0}_{T=T_{\rm f}},
\end{equation}
where $\alpha$ is defined as $\alpha=2\kappa/\lambda-3$, and $N_0=\left.dN/dy\right|_{y=0}$ stands for the rapidity density at midrapidity: 
\begin{equation}
    N_0=\frac{g A_{\rm T}}{\left(2\pi \hbar\right)^3}\frac{\tau_0}{\tau_{\rm R}}\frac{T_0^{\alpha+3}}{\alpha} \left[\frac{1}{T_{\rm f}^{\alpha}}-\frac{1}{T_{0}^{\alpha}}\right]\frac{3\pi^{3/2}\kappa}{2\lambda}\left(\frac{2\pi \kappa}{\lambda^2\left(2\kappa-1\right)-\lambda\left(\kappa-1\right)}\right)^{1/2}.
\end{equation}
The expression of $N_0$ is exactly calculated according to Eq.~\eqref{eq:N0_integral}. The initial and freeze-out temperature in the center of the fireball are denoted by $T_0$ and $T_{\rm f}$ correspondingly, while $A_{\rm T}$ stands for the average surface of the transverse plane. In Eq.~\eqref{eq:thermal_radiation_result}, the upper incomplete $\Gamma$-function is defined as:
\begin{equation}\label{eq:Gamma}
    \Gamma\left(s,x\right) = \int\limits_{x}^{\infty} t^{s-1} \exp\left(-t\right) dt.
\end{equation}
Note that Eq.~\eqref{eq:thermal_radiation_result} depends on $\kappa$ and $\lambda$ only through the parameter $\alpha$ and the midrapidity density $N_0$. It is important to note, that the value of $\alpha$ does not determine whether the flow is boost-invariant or not, since $\alpha$ depends on the ratio of $\kappa$ to $\lambda$. Therefore, for any value of $\alpha$, the flow can be boost-invariant or not depending on the equation of state. However, the pseudorapidity distributions measured in PHOBOS $Au+Au$ at $\sqrt{s_{NN}}=20-200$ GeV collisions clearly contradict the flat distribution predicted by the Hwa-Bjorken solution, so that the flow in these reactions is not boost-invariant.~\cite{PHOBOS:2004zne,PHOBOS:2010eyu} Although $N_0$ is also dependent on $\kappa$ and $\lambda$, it is not sensitive to changes in the value of these parameters due to the following reason: varying the parameters $\kappa$ and $\lambda$ can be compensated by varying the term $A_{\rm T}/\tau_{\rm R}$ with keeping the value of $N_0$ unchanged.

With the introduction of $\alpha$ and $N_0$, the thermal photon spectrum can be determined via four parameters instead of the original seven ($\tau_{\rm R}$, $A_{\rm T}$, $\lambda$, $\kappa$, $T_0$, $T_{\rm f}$, $\tau_0$ $\rightarrow$ $\alpha$, $N_0$, $T_0$, $T_{\rm f}$), which is a nice manifestation of the scaling behaviour of hydro\-dynamics. This property of hydrodynamics is particularly useful in cases where several not measurable parameters can be merged into one observable quantity.

Although I have found this scaling behaviour of the thermal spectrum only for systems with low acceleration, it has been shown in Refs.~\cite{Csorgo:2018pxh,Csorgo:2018fbz,Kasza:2018qah} through fits on pseudorapidity distributions that the acceleration of the expansion at RHIC and LHC energies is small. Furthermore, it was recently shown in Ref.~\cite{Kasza:2023fqb}, that the rapidity distribution data of different experiments collapse into a single curve by introducing a new scaling function which was derived for low accelerations ($\lambda-1 \ll 1$). These results suggest that the rate of acceleration is small not only in $Au+Au$ at 200 GeV reactions, but also in many other high-energy collisions. Hence, it can be assumed that the hydrodynamic scaling appears in the thermal spectrum of high-energy collisions.

\begin{figure}[h!]
    \includegraphics[scale=0.28]{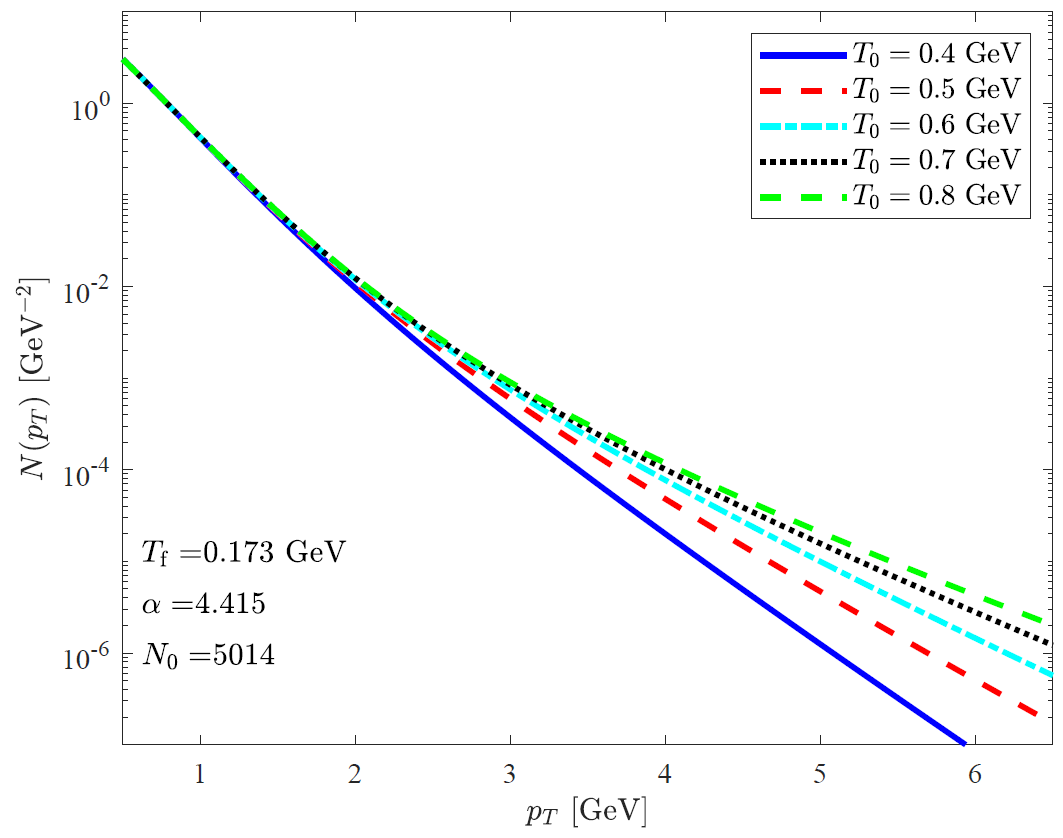}
    \includegraphics[scale=0.28]{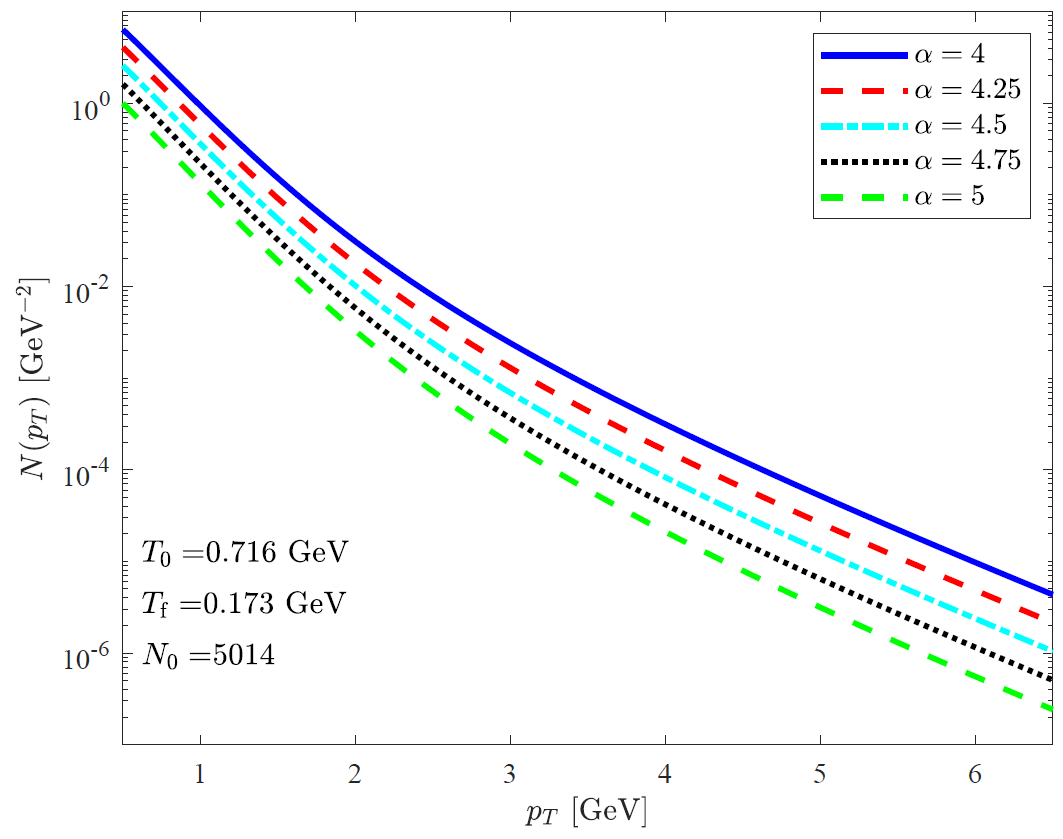}
    \caption{Left panel: Eq.~\eqref{eq:thermal_radiation_result} is plotted with fixed $N_0$, $T_{\rm f}$ and $\alpha$, but for five different values of $T_0$. This plot clearly demonstrates that the initial temperature determines the tail of the thermal photon spectrum. Right panel: Eq.~\eqref{eq:thermal_radiation_result} is drawn with fixed $N_0$, $T_{\rm f}$ and $T_0$, but for five different values of $\alpha$. This plot illustrates the hydrodynamic scaling in the thermal spectrum: each curve is described by a specific value of $\alpha$, but to each case, a number of different pairs of $\kappa$ and $\lambda$ can be assigned.}
    \label{fig:variation}
\end{figure}

Figure~\ref{fig:variation} shows two important analytical properties of Eq.~\eqref{eq:thermal_radiation_result}. In the left panel, I have plotted the thermal radiation for five different values of the initial temperature, while fixing the values of the other parameters ($N_0$, $T_{\rm f}$, $\alpha$). This plot illustrates the necessity of data points in the intermediate $p_{\rm T}$ regime to precisely determine the initial temperature. The right panel shows the manifestation of the hydrodynamic scaling of the thermal photon radiation: each curve is associated with a fixed $\kappa/\lambda$ ratio (or with a fixed $\alpha$), but the values of the parameters $\kappa$ and $\lambda$ are not fixed per se. In the same plot, the other parameters ($N_0$, $T_0$, $T_{\rm f}$) are not varied. In Fig.~\ref{fig:variation}, $N(p_{\rm T})$ denotes the double differential spectrum at $y=0$.

\section{Comparison to PHENIX $Au+Au$ Data}\label{sec:fit}

In this section, I present the fit of Eq.~\eqref{eq:thermal_radiation_result} to the nonprompt direct photon spectrum measured by the PHENIX collaboration in $Au+Au$ at $\sqrt{s_{NN}}=$200 GeV collisions with 0-20\% centrality.~\cite{PHENIX:2022rsx} This dataset is a great candidate to test my model for three reasons. First and most importantly, the prompt photon contribution was determined by $N_{\rm coll}$ (number of binary collisions) scaled $p+p$ fit, then it was subtracted from the direct photon spectrum in Ref.~\cite{PHENIX:2022rsx}. The second reason is that there are several data points in the $p_{\rm T} > 4$ GeV range, which allows a more precise determination of $T_0$. The third reason is that the solution of Sec.~\ref{sec:CKCJ} has already been successfully applied to describe hadronic observables in $Au+Au$ at $\sqrt{s_{NN}}=$200 GeV collisions with 0-30\% centrality.~\cite{Kasza:2018qah} Worth mentioning, that the initial energy density of this system has also been estimated by the solution of Sec.~\ref{sec:CKCJ}, and an almost order-of-magnitude correction to the Bjorken estimate has been found.~\cite{Kasza:2018qah}

\begin{figure}[h!]
    \centering
    \includegraphics[scale=0.45]{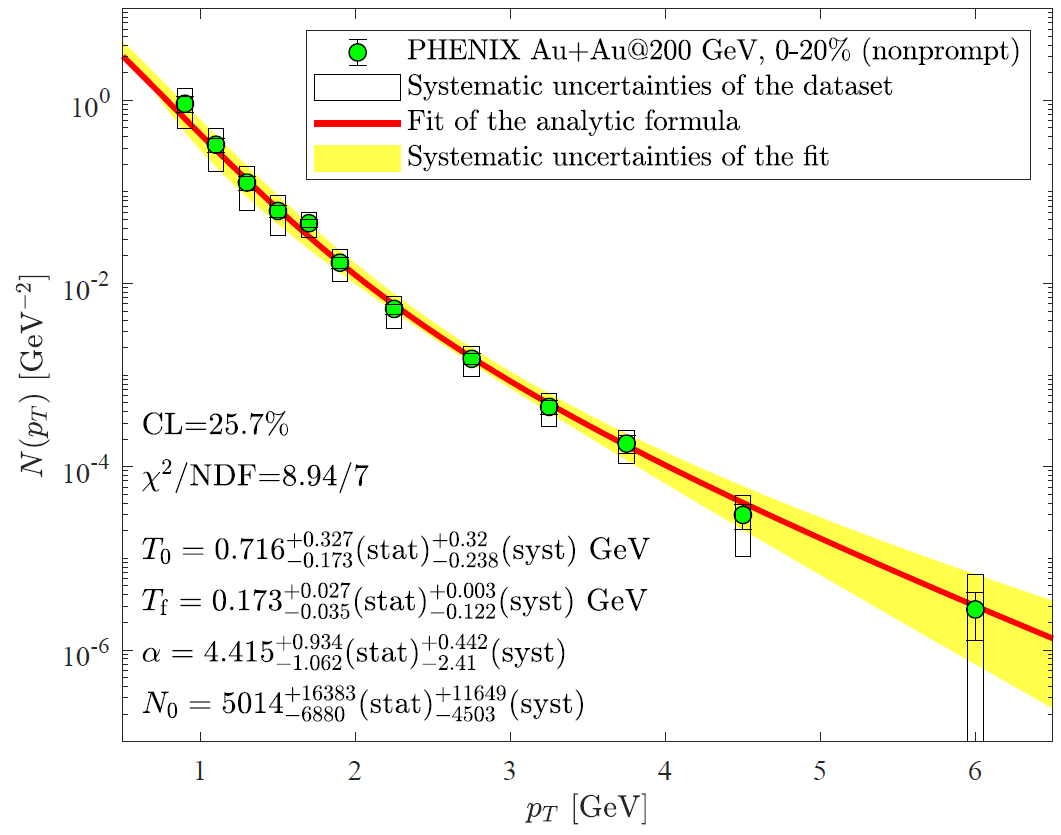}
    \caption{The fit of Eq.~\eqref{eq:thermal_radiation_result} to the nonprompt direct photon spectrum measured by the PHENIX collaboration in $Au+Au$ at $\sqrt{s_{NN}}=$200 GeV collisions with 0-20\% centrality.~\cite{PHENIX:2022rsx}}
    \label{fig:fit}
\end{figure}

The fitting of Eq.~\eqref{eq:thermal_radiation_result} to the data is shown in Fig.~\ref{fig:fit} with red line, which indicates that the new analytical formula of the thermal radiation describes the data with an acceptable confidence level. The yellow bar illustrates the systematic uncertainty of the fit. Physically realistic values are obtained for the initial temperature and the freeze-out temperature as well, and the statistical errors and systematic uncertainties of these parameters cover reasonable ranges. The value of $N_0$ is not constrained by the data, which is well reflected by the huge errors of $N_0$.

It is important to note, that Fig~\ref{fig:fit} does not include the last published data point, which lies in the $p_{\rm T}$ bin of 7 to 10 GeV.~\cite{PHENIX:2022rsx} This is because the last data point has negative value, which is not considered physical. As a result, this data point does not have a lower error, only an upper error, which is quite large. Due to that, if I fit Eq.~\eqref{eq:thermal_radiation_result} to the data that includes the last data point, the change in the parameter values shown in Fig.~\ref{fig:fit} is negligible.  

\section{Conclusions}
In this paper, I derived a new analytical formula from the 1+1 dimensional perfect fluid solution summarized in Sec.~\ref{sec:CKCJ} to describe the thermal component of the direct photon spectrum. I compared this formula with the nonprompt direct photon spectrum measured in PHENIX $Au+Au$ at $\sqrt{s_{NN}}=200$ GeV collisions with 0-20\% centrality.~\cite{PHENIX:2022rsx} The fit shows that the new formula is in agreement with the measurement, despite the fact that the solution of Ref.~\cite{Csorgo:2018pxh} needs generalization in several aspects. One of them is the 1+1 dimensional nature of the solution: the consequence of this is that although the thermal spectrum is embedded in 1+3 dimensions, the effect of radial flow does not appear in Eq.~\eqref{eq:thermal_radiation_result}, so further generalizations or corrections related to this problem are justified. Another deficiency of the solution presented in Sec.~\ref{sec:CKCJ} is the lack of viscosity, i.e. it is only suitable for describing perfect fluids. Thus, viscous effects are not taken into account in Eq~\eqref{eq:thermal_radiation_result}. However, it is certainly an interesting result that no viscosity corrections are needed in my model to obtain an acceptable description of the nonprompt direct photon spectrum data of Ref.~\cite{PHENIX:2022rsx} in the 0-20\% centrality class.

I have shown, that Eq~\eqref{eq:thermal_radiation_result} predicts the scaling behaviour of the thermal spectrum. In other words, the values of the rate of acceleration $\lambda$ and the parameter of the equation of state $\kappa$ cannot be determined from the data, only their ratio (or the parameter $\alpha$) can be extracted.

In Sec.~\ref{sec:fit}, the initial temperature in the center of the fireball was obtained and its value is indicated on Fig.~\ref{fig:fit}. This particular result should be compared with lattice QCD results and with the Hagedorn temperature~\cite{Hagedorn:1965st}. According to lattice QCD calculations, the transition between QGP and hadronic matter occurs around $T=160$ MeV.~\cite{Borsanyi:2010cj}  Although the value of the Hagedorn temperature ($T_{\rm H}$) has a wide range in the literature (Refs.~\cite{Broniowski:2000bj,Broniowski:2004yh,Cohen:2011cr,Cleymans:2011fx} suggest a value between 141 MeV and 340 MeV), my result for $T_0$ is definitely above $T_{\rm H}$:
\begin{equation}
    T_{\rm H} \ll T_0 = 0.7^{+0.3}_{-0.2}\textnormal{(stat)}^{+0.3}_{-0.2}\textnormal{(syst)}.
\end{equation}
Note that the true value of the initial temperature is likely to be lower than $T_0$, since my result does not include the effect of radial flow. However, the authors of Ref.~\cite{Csanad:2011jq} have calculated the thermal photon spectrum from a 1+3 dimensional hydrodynamic solution. Although they have taken the radial flow effect into account, they described the expansion of the fireball by a boost-invariant Hubble-type velocity field, thus they could not notice the hydrodynamic scaling in the thermal radiation. I conjecture that the scaling behaviour of the $p_{\rm T}$ spectra of thermal photons remains valid in 1+3 dimensions, which would explain why the boost-invariant calculation of Ref.~\cite{Csanad:2011jq} provided an acceptable description of the measurements. In Ref.~\cite{Csanad:2011jq}, the initial temperature in the center of the fireball was predicted to be at least $507\pm12$ MeV. Staying within the framework of analytical hydrodynamics, more realistic estimates of the initial temperature could be obtained by calculating the thermal radiation using a fireball solution with locally accelerating velocity field, that includes also the transverse dynamics. Unfortunately, as far as I know such an analytical solution of relativistic hydrodynamics is not yet known.

In conclusion, my result together with Ref.~\cite{Csanad:2011jq} suggest the formation of QGP in $Au+Au$ collisions with 0-20\% centrality at $\sqrt{s_{NN}}=200$ GeV, since the initial temperature ($T_0$) obtained from the fit is too high for the existence of hadrons. This confirms the conclusion drawn by the PHENIX collaboration in Ref.~\cite{PHENIX:2008uif}.

\section*{Acknowledgements}
I would like to thank Tamás Csörgő, Máté Csanád and Márton Nagy for the enlightening and inspiring discussions. I am also grateful to István Szanyi for his advices. My research has been partially supported by NKFIH K-133046 and MATE KKP (2023) grants. Preprint of an article published in \href{https://www.worldscientific.com/worldscinet/ijmpa}{International Journal of Modern Physics A}, 2023, \copyright World Scientific Publishing Company.

\appendix

\section{The detailed derivation of the analytic formula for the thermal photon spectrum}

The invariant momentum distribution of thermal radiation can be obtained by integrating the source function over space and time:
\begin{equation}
     \left.\frac{d^2 N}{2\pi p_{\rm T} dp_{\rm T} dy}\right|_{y=0}= A_{\rm T} \int\limits_{\tau_0}^{\infty}d\tau\int\limits_{-\infty}^{\infty} \tau d\eta_{z} \:  S\left(x^{\mu},p^{\mu}\right),
\end{equation}
where the integrals over the transverse coordinates are already performed and the average transverse size is denoted by $A_{\rm T}$. In the source function, I use Boltzmann approximation and I take into account the $H(\tau)$ window-function, so the upper limit of the integral over $\tau$ is the freeze-out proper time $\tau_{\rm f}$:
\begin{equation}\label{eq:spectra_integrals}
    \left.\frac{d^2 N}{2\pi p_{\rm T} dp_{\rm T} dy}\right|_{y=0} \approx \frac{g}{\left(2\pi \hbar\right)^3}\frac{A_{\rm T}}{\tau_{\rm R}}\int\limits_{\tau_0}^{\tau_{\rm f}}d\tau\int\limits_{-\infty}^{\infty} \tau d\eta_{z} \: \frac{p_{\mu}u^{\mu}}{\cosh\left(\Omega-\eta_z\right)}\exp\left(-\frac{p^{\mu}u_{\mu}}{T(\tau,\eta_z)}\right).
\end{equation}
The freeze-out hypersurface equation is approximated by the Hwa-Bjorken limit: $\tau_{\rm f}(\eta_z)\approx \tau_{\rm f} = \textnormal{constant}$.  This approximation makes possible to calculate the spectrum analytically, and it does not cause a significant deviation, since I am working in the $\lambda\rightarrow 1$ limiting case. The $p_{\mu}u^{\mu}$ term is the energy of the photons in the co-moving system of the fluid. In the region of midrapidity ($y\approx 0$), the longitudinal component of the four-momentum can be neglected, and I take advantage of that $\lambda$ is close to 1 (so $\Omega\approx \lambda\eta_z$):
\begin{equation}
    p_{\mu}u^{\mu} \approx p_{\rm T} \cosh\left(\lambda \eta_z - y\right)\approx p_{\rm T} \cosh\left(\lambda \eta_z \right).
\end{equation}
In the $\lambda\rightarrow 1$ limit the exponent $B$ of the source function can be written as:
\begin{equation}
    B\left(\tau,\eta_z\right)=\frac{p^{\mu}u_{\mu}}{T(\tau,\eta_z)} = \frac{p_{\rm T} \cosh\left(\lambda \eta_z \right)}{T_0}\left( \frac{\tau}{\tau_0}\right)^{\frac{\lambda}{\kappa}}\left[1+\frac{\kappa-1}{\lambda-1}\sinh^2\left(\left(\lambda-1\right)\eta_z\right)\right]^{\frac{\lambda}{2\kappa}}.
\end{equation}
The integration over the coordinate-rapidity can be performed by saddle-point app\-roximation. Using thing method, I expand the exponent $B$ to the second order around the so-called saddle-point $\eta_z^{\rm S}$. The saddle-point is defined as the main emission point of the source, i.e.:
\begin{equation}
    \left.\frac{\partial B}{\partial\eta_z}\right|_{\eta_z^{\rm S}} = 0.
\end{equation}
At midrapidity ($y=0$), the saddle-point is found to be $\eta_z^{\rm S}=0$. Performing the Taylor expansion of $B$ around $\eta_z^{\rm S}=0$, one can obtain the following:
\begin{equation}
    B\approx \frac{p_{\rm T}}{T_0}\left( \frac{\tau}{\tau_0}\right)^{\frac{\lambda}{\kappa}}\left[1 + \left(\frac{\lambda^2}{2} + \frac{\lambda\left(\lambda-1\right)\left(\kappa-1\right)}{2\kappa}\right)\eta_z^2\right].
\end{equation}
The integrand of Eq.~\eqref{eq:spectra_integrals} is a product of a Gaussian which has a sharp maximum peak at $\eta_z^{S}$ and a function which changes smoothly around $\eta_z^{\rm S}$. In this approximation the smooth function is treated as a constant equal to the value of this smooth function at $\eta_z^{\rm S}=0$:
\begin{equation}\label{eq:gaussian_integral}
\begin{aligned}
    \left.\frac{d^2 N}{2\pi p_{\rm T} dp_{\rm T} dy}\right|_{y=0} &\approx \frac{g}{\left(2\pi \hbar\right)^3}\frac{A_{\rm T}}{\tau_{\rm R}}\int\limits_{\tau_0}^{\tau_{\rm f}}d\tau \left.\frac{\tau \: p_{\mu}u^{\mu}}{\cosh\left((\lambda-1)\eta_z\right)}\right|_{\eta_z^{\rm S}}\exp\left(\frac{p_{\rm T}}{T_0}\left(\frac{\tau}{\tau_0}\right)^{\frac{\lambda}{\kappa}}\right)\cdot \\
    &\cdot\int\limits_{-\infty}^{\infty}  d\eta_{z} \: \exp\left(-\frac{\eta_z^2}{2\Delta \eta_z^2}\right),
\end{aligned}
\end{equation}
where the width of the Gaussian is:
\begin{equation}
    \frac{1}{\Delta \eta_z^2} = \frac{p_{\rm T}}{T_0}\left( \frac{\tau}{\tau_0}\right)^{\frac{\lambda}{\kappa}}\left(\lambda^2 + \frac{\lambda\left(\lambda-1\right)\left(\kappa-1\right)}{\kappa}\right).
\end{equation}
Performing the Gaussian integral and substituting the value of $\eta_z^{\rm S}$, the following result is obtained, which needs to be integrated over the longitudinal proper time:
\begin{equation}
\begin{aligned}
    \left.\frac{d^2 N}{2\pi p_{\rm T} dp_{\rm T} dy}\right|_{y=0} &\approx \frac{g}{\left(2\pi \hbar\right)^3}\frac{A_{\rm T}}{\tau_{\rm R}}\left(\frac{2\pi\kappa\:T_0 \:\tau_0^{\frac{\lambda}{\kappa}}p_T}{\lambda^2\kappa + \lambda\left(\lambda-1\right)\left(\kappa-1\right) }\right)^{1/2} \cdot\\
    &\cdot \int\limits_{\tau_0}^{\tau_{\rm f}}d\tau \: \tau^{1-\frac{\lambda}{2\kappa}} \exp\left(\frac{p_{\rm T}}{T_0}\left(\frac{\tau}{\tau_0}\right)^{\frac{\lambda}{\kappa}}\right).
\end{aligned}
\end{equation}
After performing the integral over $\tau$, one can obtain the $p_{\rm T}$ distribution of thermal photons:
\begin{equation}
\begin{aligned}
    \left.\frac{d^2 N}{2\pi p_{\rm T} dp_{\rm T} dy}\right|_{y=0} &\approx \frac{g}{\left(2\pi \hbar\right)^3}\frac{A_{\rm T}}{\tau_{\rm R}}\frac{\tau_0^2\kappa}{\lambda}\left(\frac{2\pi\kappa}{\lambda^2\kappa + \lambda\left(\lambda-1\right)\left(\kappa-1\right) }\right)^{1/2} T_0^{\frac{2\kappa}{\lambda}}\cdot \\
    &\cdot p_{\rm T}^{1-\frac{2\kappa}{\lambda}}\left[\Gamma\left(\frac{2\kappa}{\lambda}-\frac{1}{2},\frac{p_T}{T_0}\right)-\Gamma\left(\frac{2\kappa}{\lambda}-\frac{1}{2},\frac{p_T}{T_0}\left(\frac{\tau_{\rm f}}{\tau_0}\right)^{\frac{\lambda}{\kappa}}\right)\right].
\end{aligned}
\end{equation}
The upper incomplete $\Gamma$ function is defined in Eq.~\eqref{eq:Gamma}. If the following notations are introduced:
\begin{align}
    \alpha&=\frac{2\kappa}{\lambda}-3,\\
    T_{\rm f} &= T_0 \left(\frac{\tau_0}{\tau_{\rm f}}\right)^{\frac{\lambda}{\kappa}},
\end{align}
and the midrapidity density is calculated as follows:
\begin{equation}\label{eq:N0_integral}
    \left.\frac{dN}{dy}\right|_{y=0}=\int\limits_{0}^{\infty} \left.\frac{d^2 N}{dp_{\rm T} dy}\right|_{y=0} dp_{\rm T}, 
\end{equation}
the result can be written in the same form as Eq.~\eqref{eq:thermal_radiation_result} in Sec.~\ref{sec:analytic_thermal_radiation}.

\printbibliography

\end{document}